**Corrected Version**

# "Secular Light Curves of NEAs
2201 Oljato, 3200 Phaethon, 99942 Apophis,
162173 Ryugu, 495848, and 6063 Jason"

**I. Ferrín, A. Arcila, M. Saldarriaga**
**Solar, Earth and Planetary Physics Group**
**Computational Physics and Astrophysics Group**
**Institute of Physics**
**Faculty of Exact and Natural Sciences**
**University of Antioquia**
**Medellín, Colombia**
**ignacio.ferrin@udea.edu.co**

**Abstract**

Using the Secular Light Curve (SLC) formalism (Ferrín, 2010a; PSS, 58, 365-391), we searched for cometary activity of six NEAS: **2201 Oljato.** The SLC exhibits significant cometary activity and evidence of an eclipse. **3200 Phaethon.** Is the parent of the Geminid meteor stream and exhibited a faint tail at perihelion in several oppositions. The SLC fails to show cometary activity using a 1995 to 2017 data set, suggesting that we are in the presence of a dormant cometary nucleus. **99942 Apophis.** This SLC does not shows evidence of cometary activity, however it shows evidence of an eclipse. The evidence of a double nature is confirmed with Goldstone and Arecibo radar observations. **162173 Ryugu.** This NEA is the target of the Hayabusa 2 spacecraft mission. The SLC exhibits an enhancement consistent with cometary activity. It may be possible to confirm or deny this activity by the end of the mission. **495848= 2002 QD7.** This NEA shows deviations from a flat distribution near perihelion in 2007 and 2017. However our own observations in 2018 show a bare nucleus, from which we obtain an absolute nuclear magnitude $m_V$(nucleus) = 18.32±0.03 (using $p_V$=0.04). The equivalent diameter is D=1.43±0.10 km. **6063 Jason.** The SLC of this NEA exhibits low level cometary activity centred at perihelion. The fact that a cursory search for low level active comets among NEAs has uncovered four positive objects out of 6, hints that many more faint comets should exist as yet undetected.

**Key words: minor planets, comets**

**1. Introduction**

Up to 2016, 19 active asteroids showing cometary activity have been discovered inside the main belt (Jewitt et al., 2015), and the number has increased to 26 as of 2018, with a rate of ~1.2 per year. These objects were discovered at the telescope searching for asteroids with coma or tails or during sky surveys. This method consumes large amounts of telescope time (Hsieh et al., 2015b; Gilbert and Wiegert, 2009; Sonnett et al., 2011; Waszczak et al., 2013), and its is not capable of detecting low level active comets that do not exhibit a coma or a tail.

It was Opik (1963) the first to suggest that Apollo asteroids were the degassed remains of comets. And Davies (1986) suggested that 3200 Phaethon and 2201 Oljato could be extinct comets. 3200 Phaethon and 2201 Oljato have never shown cometary activity in spite of the fact that both have circumstantial evidence of having some cometary characteristics (described below). This justifies their study since they could be transition objects (asteroid → comet or comet → asteroid). Since they have never exhibited a coma or a tail (2201 for 70 years, 3200 for 34 years, 99942 for 12 years), the traditional coma/tail method of confirmation fails in their case, and some other form has to be devised.

In this work we will explore the possibility of detecting activity on 6 NEAS that have been suggested or are suspected of being active, using the Secular Light Curve (SLC) formalism (Ferrín, 2010a). The objects were selected based on their number of measurements, on being nice and representative or on their current scientific interest. However a complete and deep search was not conducted, suggesting that many active NEAs may still be hiding in the NEA population.

**2. Materials and Methods**
**2.1 Definition of absolute magnitude**

We have found that the IAU definition of absolute magnitude of an asteroid (Muinonen et al., 2010), cannot be applied to our work because the IAU definition includes the opposition effect that produces false enhancements in our plots, and thus false positives.

If the object is a comet, it has to exhibit a SLC characteristic of a comet. The SLC formalism was developed in a series of papers (Ferrín, 2005a; 2005b; 2006; 2007; 2008; 2009; 2010a; 2010b; 2010c; et al., 2012; et al., 2013; et al., 2017) and was applied to comets. The idea is to apply the same method to asteroids, calculating the absolute magnitude of the object from the equation

$$m_v(1,1,0) = m_v(\Delta, R, \alpha) - 5 \, Log \, \Delta.R - \beta. \, \alpha - \text{opposition effect} - \text{aspect Angle} \qquad (1)$$

where $m_v(\Delta, R, \alpha)$ is the observed magnitude in the V band, $m_v(1,1,0)$ is the absolute magnitude, $\Delta$ is the Sun-Earth distance, R the Sun-asteroid distance, $\beta$ is the phase coefficient, and $\alpha$ the phase angle, and to plot $m_v(1,1,0)$ vs $t-T_q$, where t is time and $T_q$ the perihelion date. The last two terms in the equation mean that the opposition and aspect angle effects have to be subtracted in some way to be specified later on. A flat distribution should be found if the object is inactive, since the absolute magnitude of an asteroid has to be independent of position in the orbit by definition (Figures 1 and 2). Since typically this involves several oppositions (~20 years and 4-15 oppositions for our plots), we call it Secular Light Curve or SLC for short (a term coined by Beatrice Mueller of the Planetary Science Institute). The SLC must not be confused with the rotational light curve that involves much shorter time periods (hours or days).

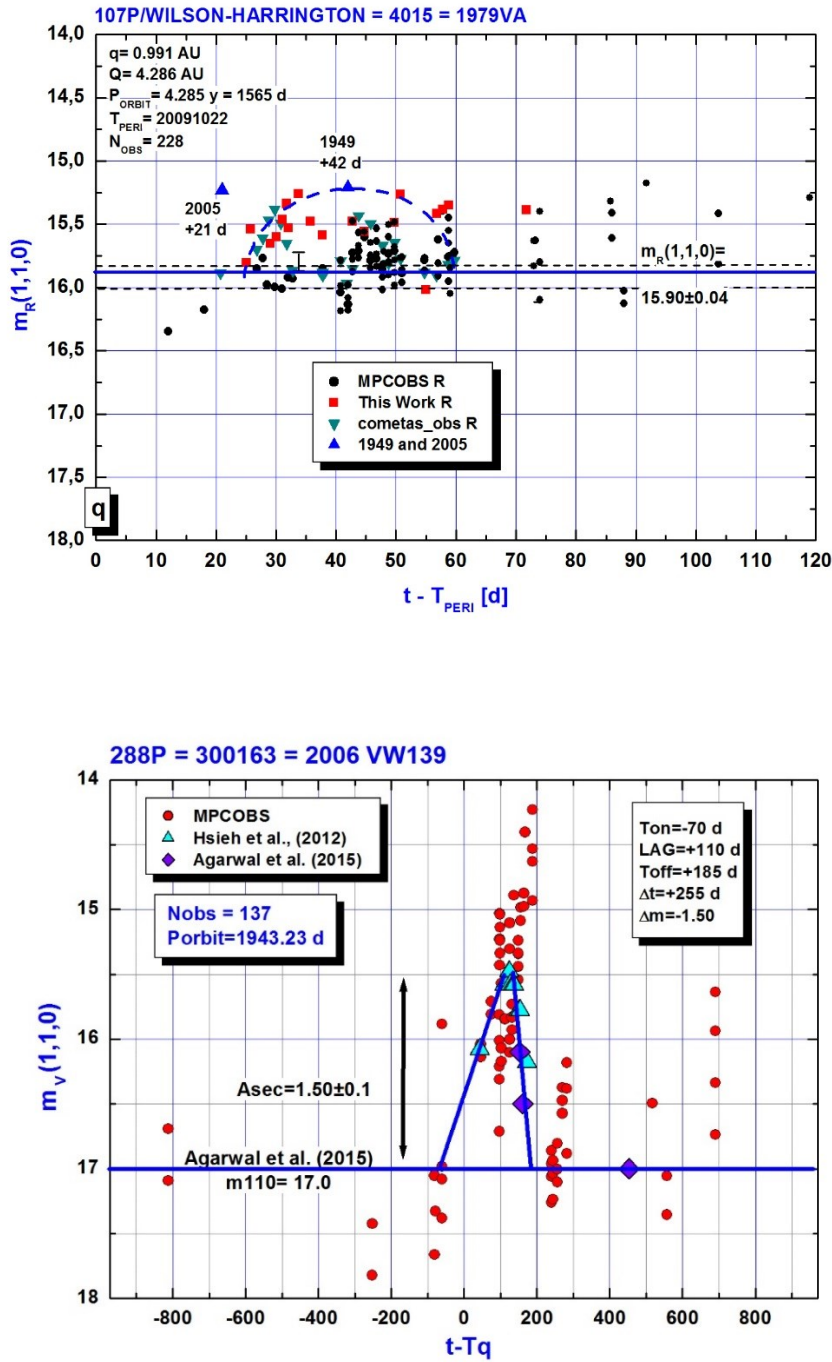

**Figure 1.** SLC of two bona fide comets, 107P/Wilson-Harrington and 288P=300163. We find for 107P (A$_{SEC}$, Δt) = (~-0.70mag, ~35d), and for 288P (A$_{SEC}$, Δt) = (~-1.50 mag, ~260 d). Cometas_obs lists magnitude observations measured with multi-apertures (Castellanos, 2018). N$_{OBS}$ is the number of observations, P$_{ORBIT}$ the orbital period.

This definition must be qualified. The definition is valid if (a) the object is spherical; (b) the polar axis is perpendicular to the ecliptic; (c) the aspect angle (polar axis vs perpendicular to the orbit) is small. Otherwise the aspect angle produces a wavy line on the SLC (Figure 3), with two maxima and two minima that is easily recognized in our plots (Ferrín et al., 2017). None of the objects discussed in this work exhibit such a signature, thus concluding that the aspect angle does not influence the results arrived at in this manuscript.

Concerning the opposition effect it must be emphasized that it is constrained to phase angles $\alpha < 5.5°$, and it is not included in our definition of absolute magnitude as it is in the IAU interpretation of the phase plot (Muinonen et al., 2010), because it produces false positives in our plots. In our formalism the Earth-object and Sun-object distances, the phase angle, the aspect angle and the opposition effect are geometrical effect that have to be removed from the absolute magnitude definition. What is left is the true activity of the objects. If this is done, our assertion that the absolute magnitude is independent of the location on the orbit is correct, and this is fulfilled in the two examples given in Figure 2 and not fulfilled in the two examples given in Figure 3 that shows the influence of the aspect angle on the SLC.

As an example of the difference in the two calculations, Ishiguro et al. (2014) find an absolute magnitude Hv=19.13±0.03 for the Hayabusa-2 Mission target 162173 Ryugu using the IAU formalism, while from their Figure 2 a value $m_V(1,1,0)$=19.45±0.03 can be read using the linear formalism that ignores the opposition effect, a non-negligible difference of ~0.32 magnitudes. In other words, in this work

$$Hv \neq m_V(1,1,0), \qquad m_V(1,1,0) \sim Hv + 0.32, \qquad Hv \sim m_V(1,1,0) - 0.32$$

On the other hand, if the object is active, the m(1,1,0) vs t-Tq diagram will show some kind of enhancement with respect to the nucleus line, like observed in comets. The distribution of observations is not flat and in many occasions it is brightest at perihelion or just past perihelion. The last possibility might be due to a thermal lag since the thermal wave takes some time after maximum surface temperature, to reach a shallow layer of ice buried below the surface.

We will use the SLC formalism to shed some light into the behavior of 6 NEA objects: 3200 Phaethon, 2201 Oljato, 99942 Apophis, 162173 Ryugu, 495848 and 6063 Jason. A resume of our numerical results appears in Table 1.

## 2.2. Definition of Envelope

In this work we adopt the *envelope of the data set* as the correct interpretation of the observed secular light curve. There are many physical effects that affect comet observations such as twilight, moon light, haze, cirrus clouds, dirty optics, old mirror coating, lack of dark adaptation, excess magnification, poor collimation, and in the case of CCDs, sky background too bright, insufficient exposure time, insufficient CCD measurement aperture, and in one case we know, the photometry tool has a maximum aperture smaller than the coma's diameter. All these factors diminish the captured photons coming from the comet, and the observer makes an error downward, toward fainter magnitudes. Thus *the envelope* is the correct interpretation of the light curve. This interpretation is confirmed by the dataset: the envelope is rather smooth and sharp, while the anti-envelope is diffuse and irregular.

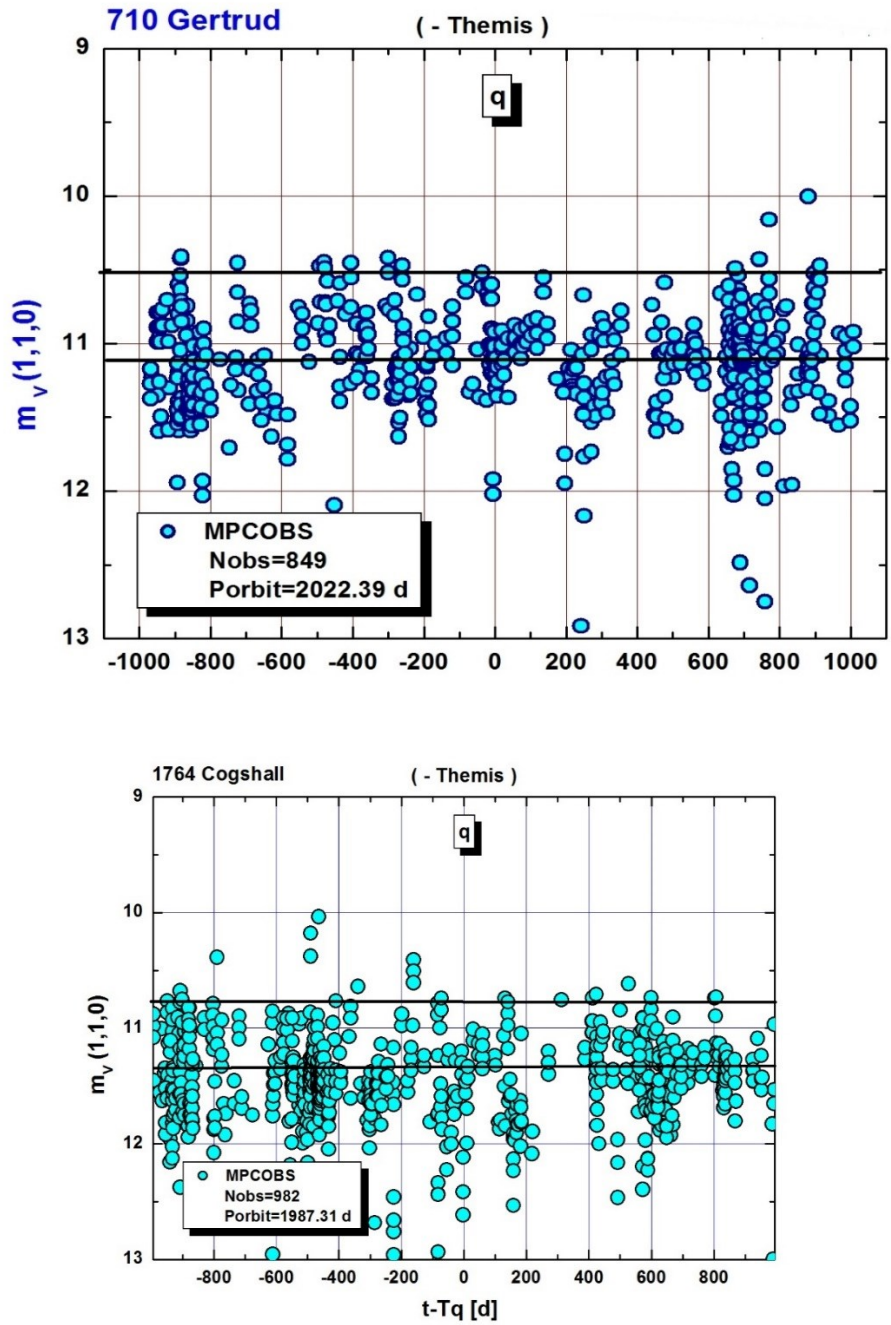

**Figure 2.** SLCs of two inactive asteroids, 710 Gertrud, and 1764 Cogshall. The distribution is quite flat with no significant events, thus $(A_{SEC}, \Delta t) = (0,0)$ for both.

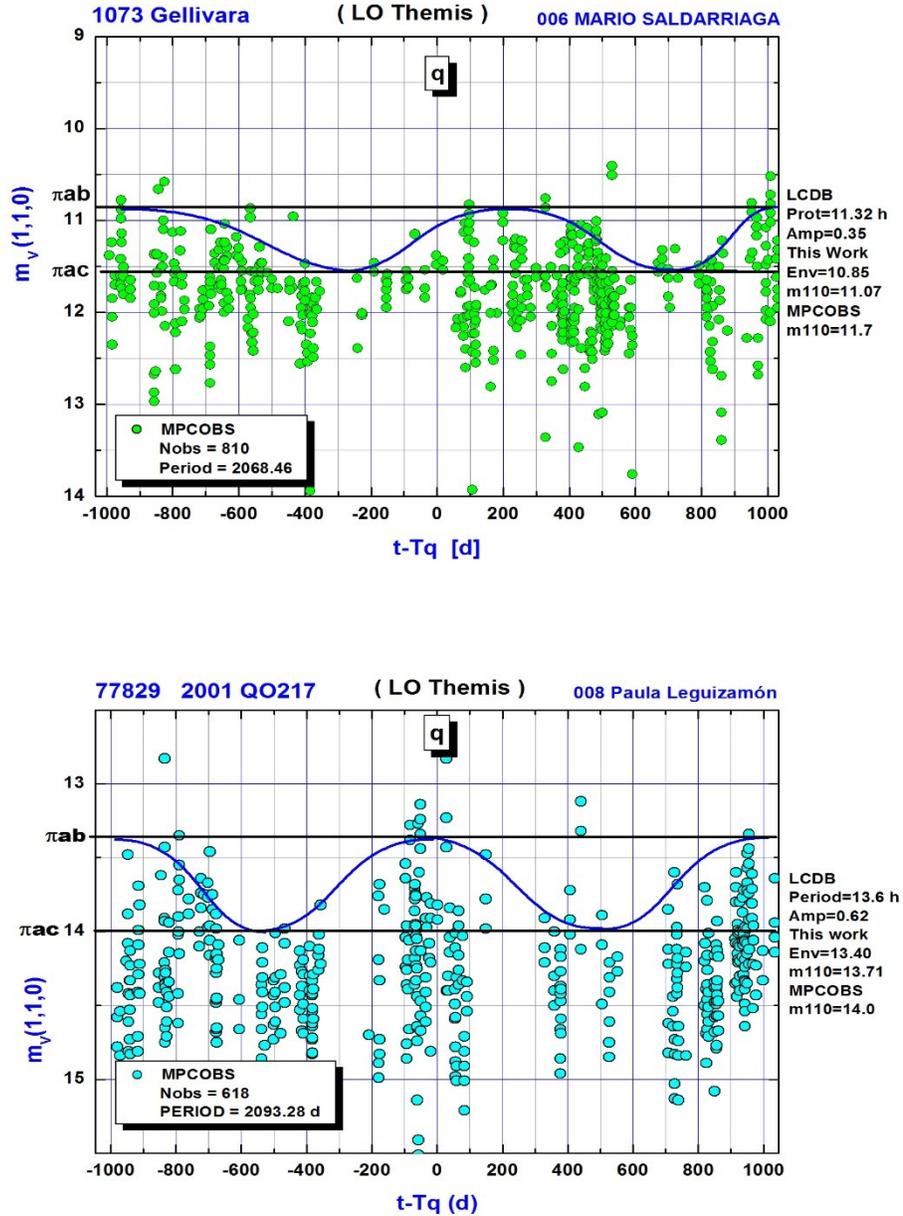

**Figure 3.** Asteroids with large obliquity (aspect angle) show a well-defined signature in their SLCs, consisting of a wavy line with two maxima and two minima (Ferrín et al., 2017). We show two cases where this signature is very strong. None of the objects presented in this work show this signature, thus the aspect angle cannot influence our interpretation of the SLCs. Notice the label on the Y-axis from which a value for the ratio b/c of a regular ellipsoid can be derived. The Asteroid Light Curve Database, LCDB, is administered by Warner et al. (2009).

The envelope represents an ideal observer, using an ideal telescope, with an ideal detector, in an ideal atmosphere.

## 2.3 Definition of dormant, extinct and low level active comet
For the sake of clarity in this investigation we will adopt the following definitions:

Dormant comet = a comet that has volatiles buried in the nucleus but there is no sublimation all around the orbit. It is indistinguishable from an asteroid. A dormant comet may return to activity if its perihelion distance were to decrease. In this case the thermal wave would penetrate deeper, activating the buried layer of ice and returning the object to active duty. These are known as a Lazarus comets (Ferrín et al., 2013)

Extinct comet = a comet that has exhausted all volatiles in the nucleus. It is indistinguishable from an asteroid.

Low level active comet = a comet that has volatiles, it is sublimating, but the amplitude of the SLC is < 3 magnitudes. It does not show a coma or a tail.

The selection of the <3 magnitudes value is explained in the next Section.

## 2.4 Definition of Threshold Coma Magnitude
It has be found observationally that comet 28P/Neujmin 1 exhibited a coma with V(observed) - $V_{NUC}$ = -3.2 mag, where $V_{NUC}$ is the nucleus magnitude. Comet 133P/Elst-Pizarro did not exhibit a coma with V- $V_{NUC}$ = -2.8 mag. Comet 2P/Encke did not exhibit a coma at aphelion with V- $V_{NUC}$ = -2.4 mag. Lastly, comet 107P/Wilson-Harrinton did not exhibit a coma with V- $V_{NUC}$ = -0.70 mag (Figure 1). Thus there seems to be an intermediate value at which the coma disappears hidden inside the seeing disk. This quantity is defined as the *Threshold Coma Magnitude* and its estimated value is TCM = ~3.0±0.2 magnitudes (Ferrín, 2012). The dependence of this value on the FWHM of the seeing disk, has not yet been quantified.

## 2.5 Comparison objects
Before going into our objects of interest, it is important to get acquainted with the SLCs of two low level active comets (107P/Wilson-Harrington and 288P/300163) (Figure 1), two non-active asteroids, 710 Gertrud and 1764 Cogshell (Figure 2), and two asteroids than exhibit large obliquity and aspect angle (Figure 3). Any confirmed alteration of the SLC is ascribed to cometary activity especially if it is centred or past perihelion. An alteration is defined as an enhancement (thus a negative correction), or a bump, of amplitude $A_{SEC}$, and duration $\Delta t$ in days ($A_{SEC}, \Delta t$). Asteroids 710 Gertrud and 1764 Cogshell do not show any alteration, the distribution is flat, thus fulfilling the definition of absolute magnitude (having the same value anywhere around the orbit). Thus their parameter ($A_{SEC}, \Delta t$) =(0,0).

### 3. Results and discussion
### 3.1 2201 Oljato

According to the JPL Small-Body Database Browser and the IRAS Supplemental Catalog (Tedesco, 2002) Oljato is a NEA of absolute magnitude H=15.25, diameter D=1.8 km and geometric albedo pv=0.43. Its perihelion distance is q=0.62 AU and its orbital period $P_{ORBIT}$ = 3.21 y. The CALL database lists 2201 with a rotational period of 26 h and amplitude ~0.1 mag (Warner et al., 2009). Green (1985) lists the object as an Apollo asteroid with a 1.9±0.4 km diameter from the IRAS data.

The question of whether 2201 Oljato is a comet or an asteroid is a long-standing one (Drummond, 1982; Davies, 1986). Drummond (1982) suggest a possible cometary origin for this asteroid on the basis of a comparison to known meteor streams including the χ Orionids and the Canids. MacFadden et al. (1984) reported that 2201 has a reflectance spectrum which is distinctly non-asteroidal in character. McFadden (1990) made a re-analysis of the 1979 photometry using the Haser model, and found a production rate larger than for comet 28P/Neujmin 1. The UV activity could be explained as a result of residual outgassing, as in a comet. Their calculation indicates a plausible value for OH and CN emission within the observed range of other active comets. McFadden et al. (1993) list a number of reasons why Oljato is not expected to be a dormant or extinct comet, starting with the high albedo and its classification as an E asteroid.

Other evidence that Oljato may be a comet comes from the detection of disturbances in the interplanetary magnetic field observed by the Pioneer Venus Orbiter, correlated with the passage of Oljato near the planet (Russell et al. 1984). These disturbances are not due to Oljato itself, but to interactions of the solar wind with material in the same orbit, lagging behind the object by up to 70 days.

More recently Lai et al. (2014) have confirmed these disturbances. They interpreted these interplanetary field enhancements (IFEs) as due to material in Oljato's orbit, colliding with material in or near to Venus' orbital plane, that produces a dust-anchored structure in the interplanetary magnetic field. The material detected by Lai et al. (2014) would naturally be explained if 2201 were a comet and not an asteroid.

### 3.1.1 Our results

In order to explore the above hypothesis, we created the phase plot shown in Figure 4 , and the SLC of 2201, shown in Figure 5. The phase plot was fitted with a linear law. Four different oppositions show activity at the same place in the orbit. What is the evidence of cometary activity then?: (a) Recurrent activity at different returns is currently considered the most reliable indicator of sublimation driven activity (Hsieh et al., 2015). (b) Additionally, if activity peaks at or near perihelion then sublimation is the most probable reason of the enhancement.

The location of 2201 in the $A_{SEC}$ vs Diameter diagram is shown in Figure 22. The object lies in the graveyard of comets defined as $A_{SEC}$ <1.0 magnitudes.

### 3.2 (3200) Phaethon

According to the IRAS catalog (Tedesco et al., 2002), Phaethon is a NEA of absolute magnitude H=14.51, and geometric albedo 0.11. Its perihelion distance is only q=0.14 AU and

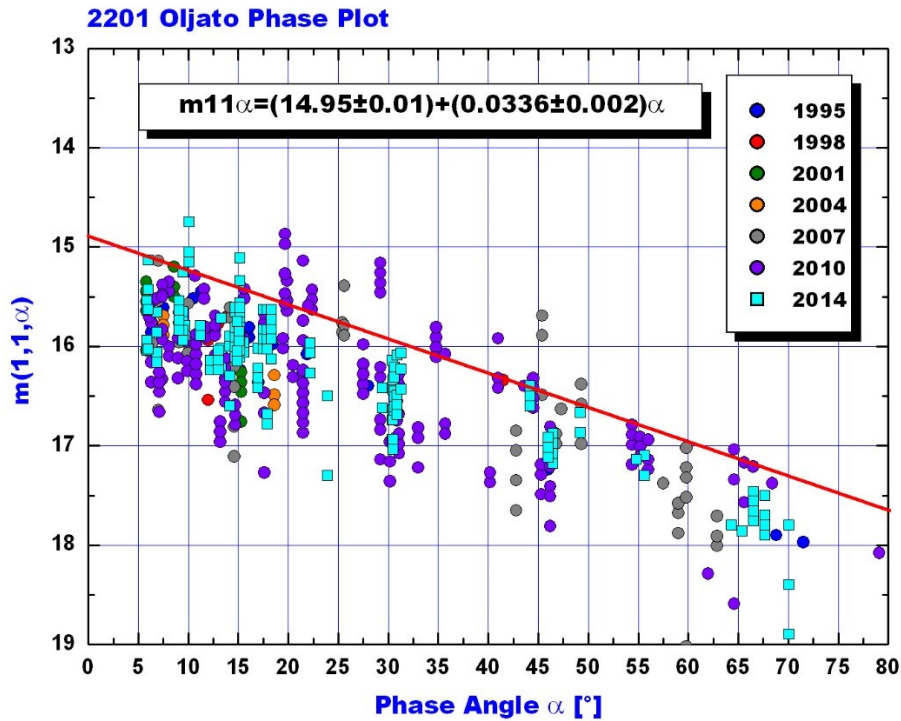

**Figure 4**. Phase plot of 2201 Oljato. The data distribution can be fitted with a linear law of slope 0.036 up to α = 70°. The best way to determine the absolute magnitude of the object is extrapolating the observed magnitudes to phase angle α =0°. We find $m_V(1,1,0)$=14.90±0.01 and β=0.035±0.02. All data points with α< 5.5° have been removed, otherwise the opposition effect would create false positives in our plots.

o o o o o o o o o o o o o o o o o o o o

its orbital period $P_{ORBIT}$ =1.49 y. The JPL Small Body Database lists 3200 with H=14.6, D=5.10 km, pv=0.1066. The CALL database (Warner et al., 2009) gives its rotational period and amplitude, $P_{ROT}$ =3.6 h and $A_{ROT}$=0.34 mag.

The issue if 3200 Phaethon is a comet or an asteroid is an old one, too. Davies (1986) suggested that it could be an extinct comet, based on the fact that 3200 Phaethon is the parent of the Geminid meteor stream (Gustafson, 1989; William and Wu, 1993).

Jewitt and Li (2010) detected activity and a tail during 7 days in May of 2009 and 4 days in 2012 (Li and Jewitt, 2013), just past perihelion (Jewitt et al., 2013). Previous attempts to detect activity by Hsieh and Jewitt (2005) and Wiegert et al. (2008) were unsuccessful. They propose thermal fracture as mechanism to explain the activity.

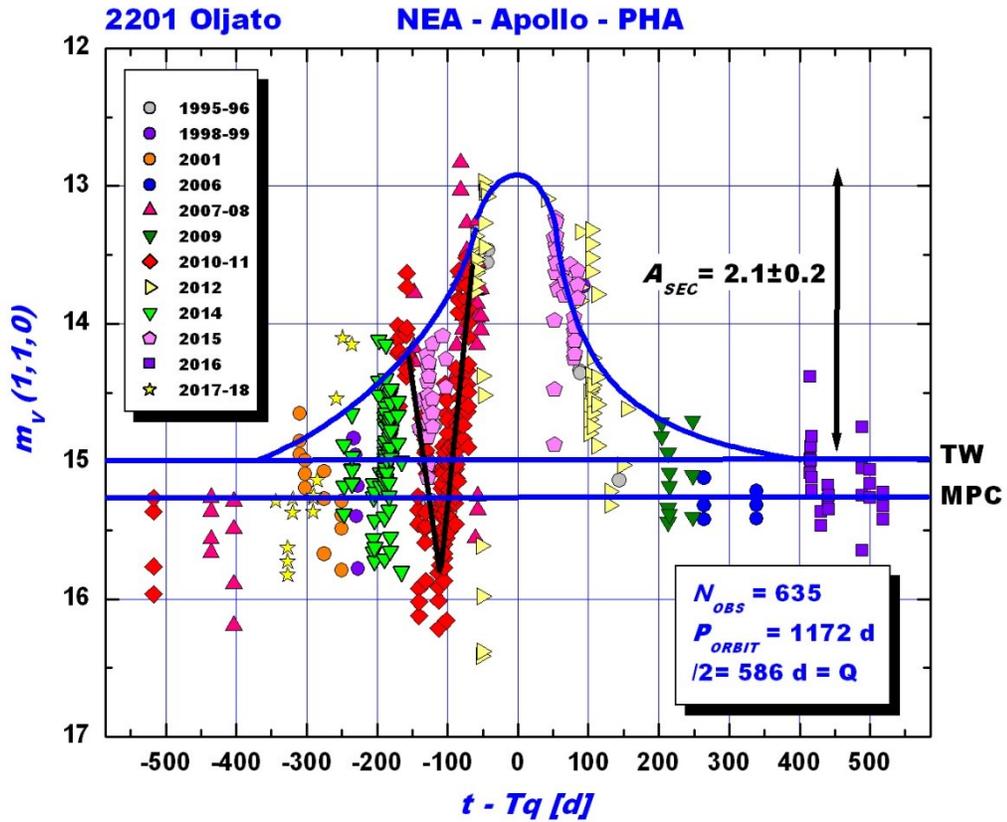

**Figure 5.** The SLC of 2201 Oljato shows enhanced activity centered at perihelion on many oppositions suggesting that 2201 is a low activity bona fide comet. Recurrent activity at different returns is currently considered the most reliable indicator of sublimation driven activity (Hsieh et al., 2015). Additionally, activity is centred at perihelion, clear indication of a active sublimation. In this way we resolve a question that has been pending for 34 years (Drummond, 1982). We conclude that this method of reduction (SLCs) is capable of detecting low activity comets. Notice that 2201 Oljato has never exhibited a coma, thus this enhancement remained undetected for more than 69 years. The reason might be that the coma of this object is very small and has remained all this time contained inside the seeing disc. The pattern cannot be induced by the rotational light curve, since the amplitude of the later is $A_{ROT}$=0.1 mag (Warner et al., 2014). We find ($A_{SEC}$,$\Delta$t) = (~-2.1,~700d). TW = This Work. This SLC shows evidence of an eclipse.

The very small activity at perihelion observed in 3200 Phaethon using the space-based STEREO solar observatory, was interpreted to arise from thermal fracture and desiccation cracking due to the temperature, rather than comet like activity from buried volatiles (Jewitt and Li, 2010).

More recently observations of (3200) Phaethon were carried out by Hui and Li (2016) using STEREO. They were able to measure the absolute V magnitude of the object as a function of phase angle (their Figure 6). They found, as previously, and enhancement of about ~2 magnitudes at perihelion. Hui and Li (2016) measured a released mass of ~ $10^4$ to $10^5$ kg, comparable to the mass ejections in 2009 and 2012. The Geminid meteor stream is a dynamically young stream, which suggests ongoing activity, yet (3200) Phaethon has exhibited no known comet-like outbursts sufficient to replenish the stream (Hughes and McBride, 1989; Jenniskens, 1994).

Boice et al. (2013) performed a detailed three-dimensional chemical modeling of (3200) Phaethon using a highly oblique pole similar to the one found in the work by Ansdell et al. (2014), in order to assess whether water ice could still exist in the core of (3200) Phaethon. They found that (3200) Phaethon is likely to contain relatively pristine volatiles in its interior despite repeated close approaches to the Sun, suggesting that this might be a dormant comet. These characteristics justify a detailed study of 3200.

### 3.2.1 Our results

Ansdell et al. (2014) have measured the phase curve of (3200) finding a best-fit model parameters of H=13.90 and G=0.06. This phase curve is plotted in our Figure 6, where we find a very similar H value of H=13.17 and the same G for the data set of the Minor Planet Center. Ansdell et al. (2014) conclude that G ~0.06 is consistent with C-type asteroids and comets. The linear fit gives m $(1,1,\alpha)= (13.70\pm0.03)+(0.035\pm0.04)* \alpha$ .

### 3.2.2 3200 Secular Light Curve (SLC)

In Figure 7 we show the SLC of (3200) Phaethon using the Minor Planet Center data base of astrometric-photometric observations of minor planets. The absolute magnitude has been calculated from Equation 1. Figure 7 shows no detection of cometary activity on asteroid 3200 Phaethon on a dataset from 1995 to 2017, a total span of 22 years. This is a surprising result after all the evidence presented above on a possible cometary origin for this object. Hughes and McBride (1989) give a total mass of the Geminid stream of ~1.6 x $10^{13}$ kg, while the total mass of the object using a density of 2000 kg/m is around 1.3 x $10^{14}$ kg. The mass of the stream is too large in comparison with the mass of the object to be caused by a collision. Thus we conclude that 3200 is an extinct or dormant comet (see definition above in Section 2.3). The possibility of being an extinct comet is ruled out by the result of Boice et al. (2013).

If 3200 Phaethon is a dormant comet, it might return to activity if its perihelion distance q were decrease. In those cases the thermal wave penetrates inside the nucleus and reactivates a deep layer of ices. In that case we would be in the presence of a Lazarus comet (Ferrín et al., 2013), defined as a dormant object that returns to activity after a decrease in q. However this possibility is negated in the near future. In Figure 8 we show the evolution of the perihelion distance of 3200 during two centuries, and the result is that q is increasing, with the consequence that the solar insolation is decreasing at a rate of -1.88% per century in the foreseeable future quenching any penetrating thermal wave.

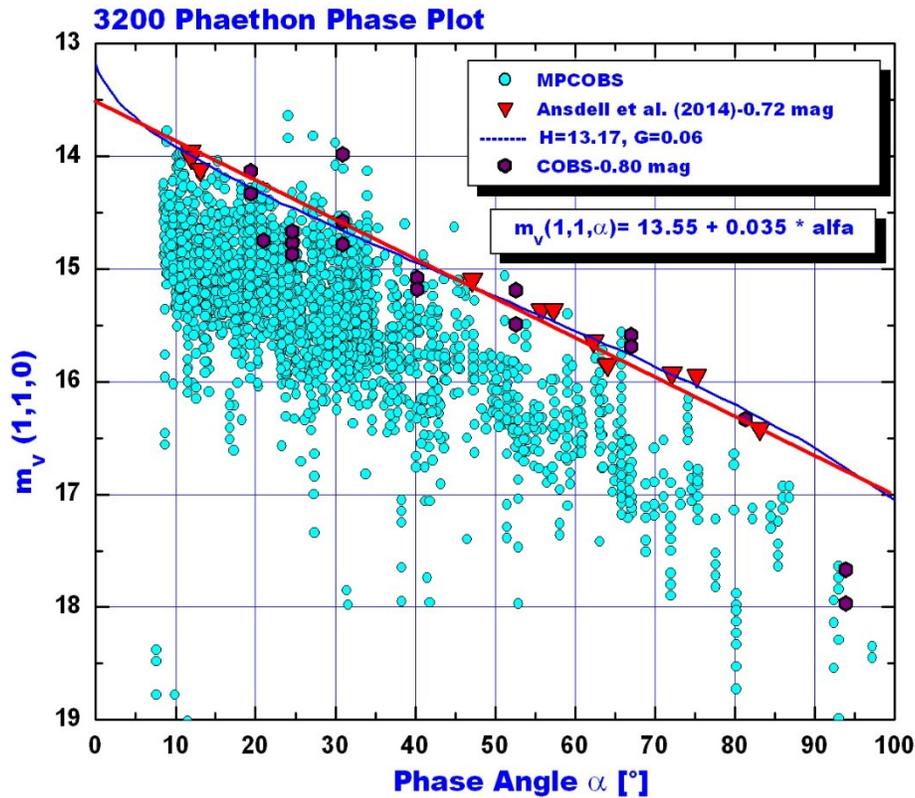

**Figure 6.** Phase diagram of (3200) Phaethon with data from the MPC and the IAU formalism phase curve of Ansdell et al. (2014) with H=13.17 and G=0.06. The plot show that the linear fit and the IAU formalism are almost identical in the range of 0° to 100°. All data points with α< 5.5° that would produce an opposition effect have been removed, since they would produce false positives. The COBS database (Zakrajsek, 2018) contains visual observations of this object.

0 0 0 0 0 0 0 0 0 0 0 0 0 0 0 0 0

### 3.3 (99942) Apophis

Apophis is a NEA and a PHA. According to the MPC as a result of its passage within 66.000 km of the Earth on 2029 April 13th, this minor planet will move from the Aten to the Apollo class (Minor Planet Circular 54567). The MPC list Apophis with a perihelion distance q=0.75 AU, orbital period $P_{ORBIT}$= 0.89y and absolute magnitude H=19.2. According to Delbo et al. (2007), it has a diameter of 270±60 m, an absolute magnitude H=19.7±0.4, and a geometric albedo pv=0.33±0.08. Licandro et al. (2015) find effective diameter D=380-393 m, pv=0.24-

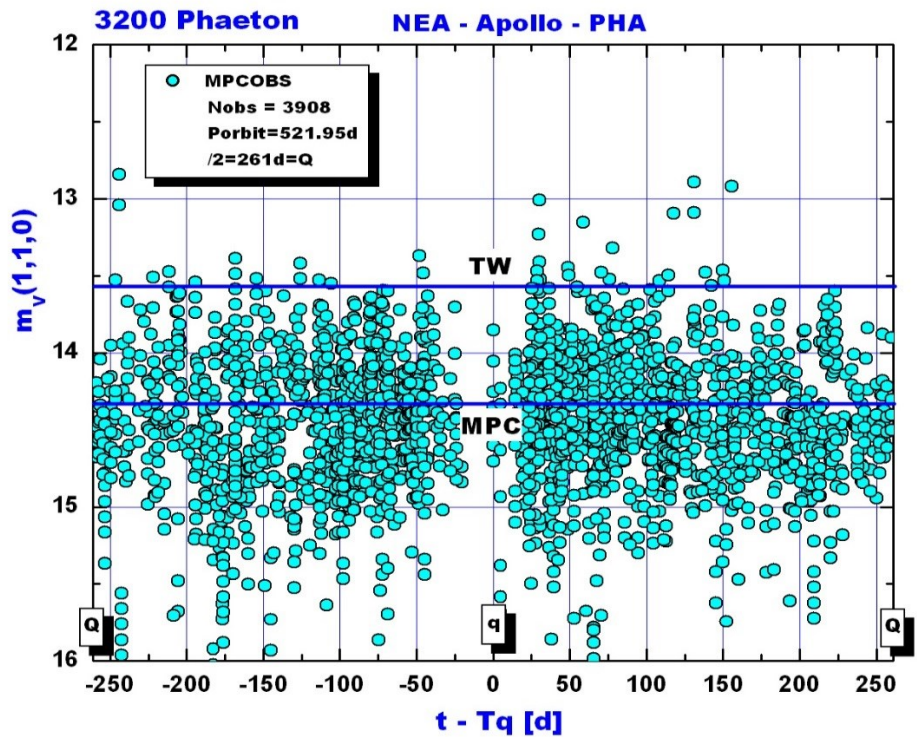

**Figure 7.** The Secular Light Curve of 3200 Phaethon. 3908 observation of the object from the MPC database obtained from 1995 to 2017, were reduced to get the absolute magnitude vs time to perihelion. The distribution is quite flat suggesting that the object is inactive all along the orbit. Combining this information with previous information, one possible explanation is that 3200 is an extinct comet. Notice the significant difference in the absolute magnitudes found in this work (TW) and that listed by the MPC.

o o o o o o o o o o o o o o o

0.33 and H=19.09±0.19. The CALL database (Warner et al., 2009) does not list a rotational period, but it lists a maximum amplitude of 1.14 mag for this object. The JPL small body database lists a $P_{ROT}$ =30.4 h. It is interesting to check if it has cometary activity and one way to do it is to create its SLC from data taken from the Minor Planet Center database.

### 3.3.1 Our results

The phase plot is shown in Figure 9 and the SLC in Figure 10. The object has a perihelion distance of q<1 AU and thus most of its time is expent inside the Earth's orbit,

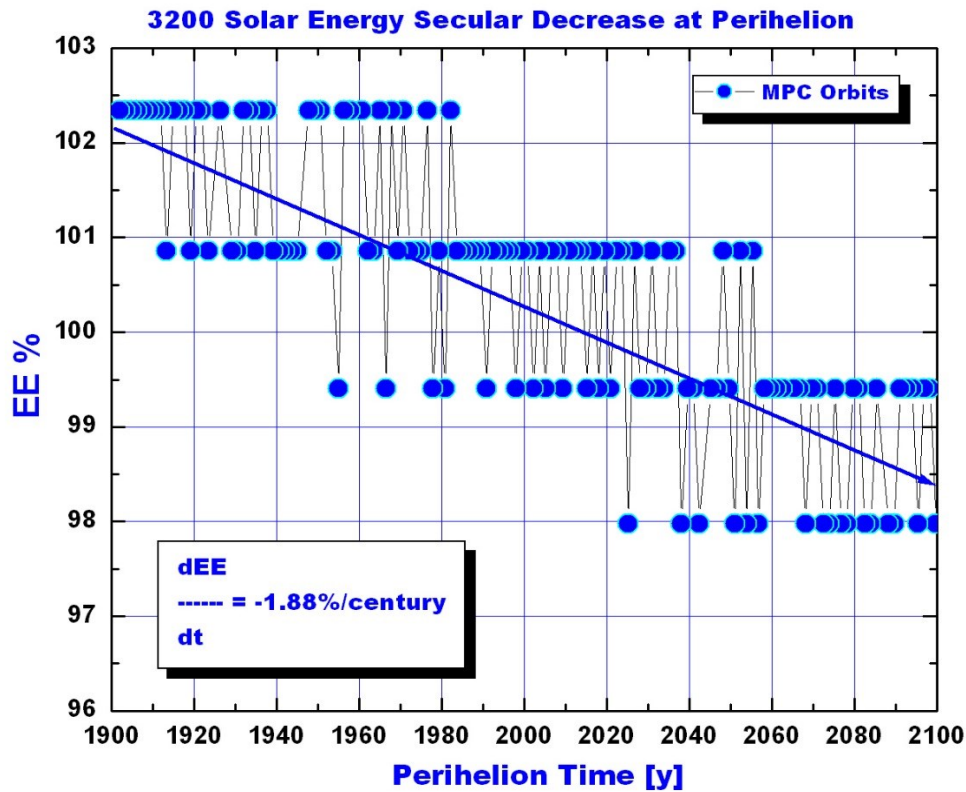

**Figure 8.** Energy received from the Sun at perihelion in an arbitrary scale with the year 2018 = 100% energy. The asteroid is receding from the Sun during these two centuries. In the next century the object will receive 1.88% less energy, thus its probability of getting re-activated by a perihelion decrease are remote.

o o o o o o o o o o o o

where observational conditions are difficult from ground based telescopes. Thus there is an observational gap at the center of the plot. The SLC exhibits no evidence of cometary activity, but it shows evidence of an eclipse, suggesting that 99942 might be a binary. The eclipse started -159±2 days and ended -121±2 days before perihelion with a depth of $\Delta m = +1.15\pm0.10$ mag.

Evidence of bifurcation in the shape of the object was uncovered by Brozovic´ et al. (2018) using Goldstone and Arecibo radar observations in 2012-2013. In their Figure 2A the image of Jan 14, run 32-64 is the one that exhibits more clearly the bifurcation because we are observing the object side-on. This image is reproduced in our Figure 11 and was rotated, and filtered with a Gaussian of width 3.5 pixels to iron out the pixelation. Mutual eclipses of this shape would produce the observed eclipse signature.

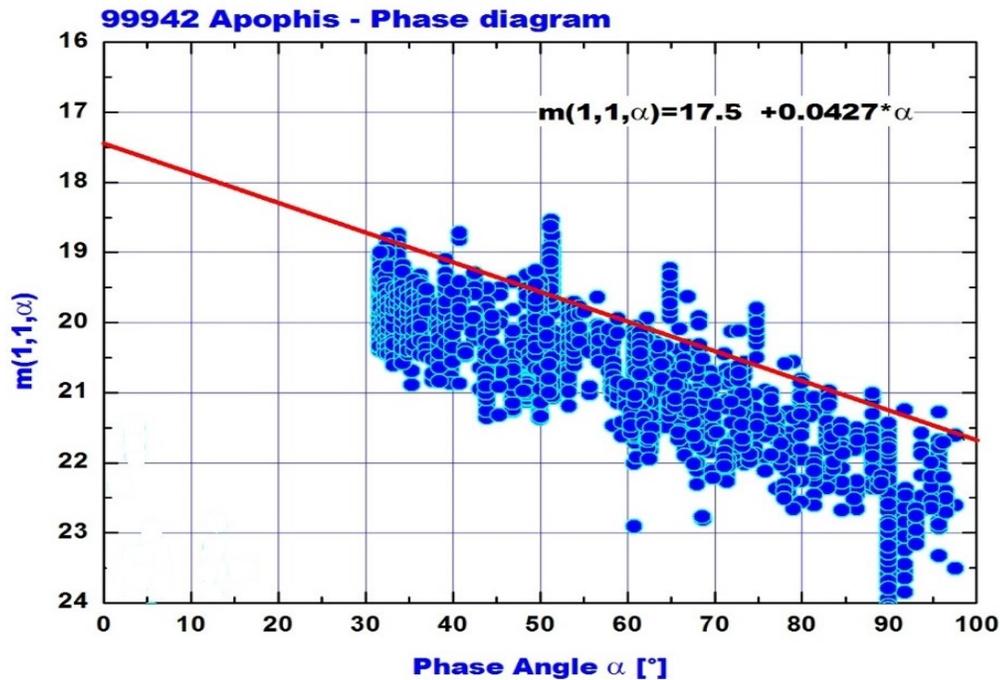

**Figure 9.** The phase plot of 99942 Apophis can be fitted with a linear behavior of absolute magnitude $m_V(1,1,0)=17.5$ and slope 0.0427. The MPC lists the absolute magnitude $H_V= 19.2$ but their own data (this data) gives quite a different result probably because we use the envelope as the interpretation of the SLC, while it seems the use the average value.

0 0 0 0 0 0 0 0 0 0 0 0 0

Based on the radar observations Brozovic' et al. (2015) derive a shape 0.43x0.30x0.26 (±0.04, ±0.03, ±0.03) km for long, intermediate and short axis, however, and surprisingly, these numbers do not fit the image in Figure 11, and the relative size of the two components was not derived either. Based on this image the shaft is about 2.2 times longer than the head, thus the head should be only ~110 m in diameter. The current estimate of the approach distance of this object to Earth in 2029 is 66.000 km, thus the object should display an angular dimension of only ~0.8 arc seconds on the approach date.

The next occasion to observe the eclipse region from ground based telescopes will be starting on 2020 04 12 when the object will be at Elongation= 76° and magnitude = 21.0, and on 2021 03 01 when the object will be at Elongation= 156° and magnitude = 15.7. However it is advisable to start observing much before those dates in case the mutual eclipse is total.

### 3.4 162173 Ryugu

This object is the target of the Hayabusa-2 Mission (Anonymous, 2017). Hayabusa-2 aims at acquiring samples and bringing them back from a C-type asteroid, to elucidate the origin

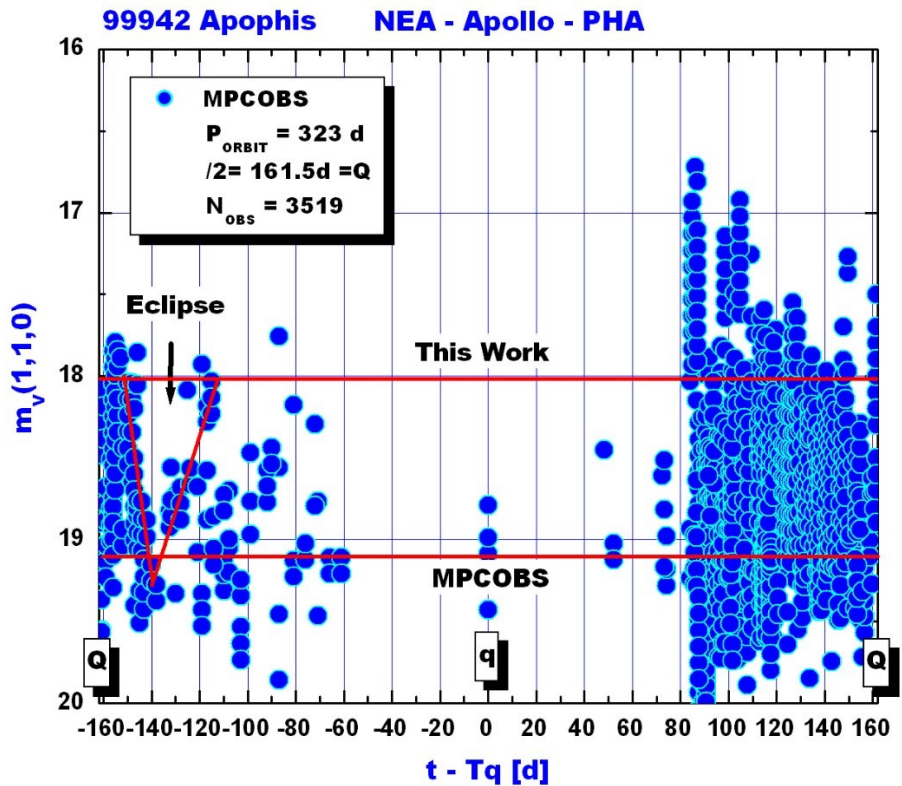

**Figure 10**. The SLC of 99942 Apophis. The perihelion of this object is interior to the Earth's orbit, thus it is difficult to observe it near the Sun. Notice the large difference in the absolute magnitude obtained in this work and the MPCOBS database. This light curve exhibits evidence of a partial eclipse lasting 32±3 days, suggesting that 99942 might be a binary. Evidence of bifurcation has been discovered by Brozovic' et al. (2018) (see next Figure). This Figure shows that the SLC formalism is good at detecting eclipses.

o o o o o o o o o o o o o o o o o

and evolution of the solar system and material for life. Hayabusa arrives at the asteroid on June-July 2018 and departs from it on December 2019.

Using a near-IR spectrum Pinilla-Alonso et al. (2013) confirm the primitive nature of this asteroid as a C-type. Muller et al. (2016) find that the asteroid has retrograde rotation with an effective diameter of 850-880 m, geometric albedo 0.044-0.050, thermal inertia 150-300 J m$^{-2}$ s$^{-0.5}$ K$^{-1}$. Ishiguro et al. (2014) find and absolute magnitude H=19.25±0.03 and colours V-R$_C$=0.37±0.03. They also fitted the phase plot with three different functions. Their linear fit gave a value for β= 0.039±0.001 while our result (Figure 12) gives β =0.036±0.002. The agreement is quite satisfactory and is within the errors. Kawakami (2009) measured colours B-

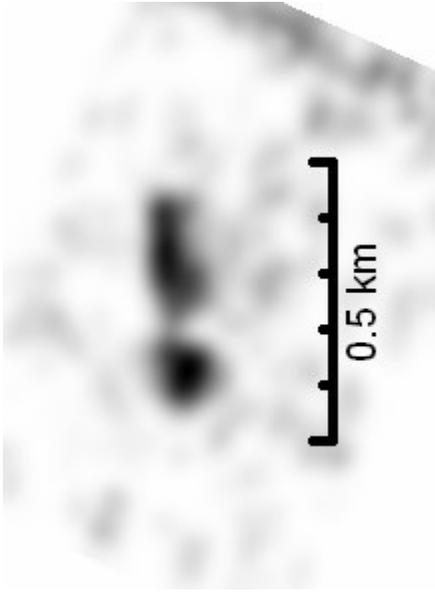

**Figure 11.** Radar image from Jan 14, runs 32-62, Figure 2A of the work by Brozovic' et al. (2018) (reproduced with permission from Elsevier), was rotated and blurred with a Gaussian of width 3.5 pixels. The result shows that 99942 is composed of two pieces in the shape of an exclamation mark (!). This shape differs substantially from the shape derived by Pravec et al. (2014) who obtained an elongated ovoid with wide and tapered ends and ratio of axes a/b= 1.44. Their shape model did not account for possible concavities or bifurcations. From this image a much larger axis ratio can be measured. A very thin bridge resembling the one observed in comet 67P/Churyumov-Gerasimenko unites the two pieces. What we believe is a mutual eclipse of this system has been detected in Figure 10. The image is reproduced in negative to show the faint detail of the bridge.

0 0 0 0 0 0 0 0 0 0

V=0.66±0.06, V-R=0.40±0.04 and V-I=0.74±0.07, from which a R-I=0.36±0.04 can be calculated.

Campins et al. (2009) and Mueller et al. (2011) find a spherical shape for the object. Muller et al. (2017) find for the ratio of the two semi-axis a/b=1.025. Thus the SLC should not show the signature of large obliquity, which is observationally confirmed.

In 2007 July Vilas (2008) found evidence of a 0.7µ absorption feature commonly seen in C-type asteroids which has been attributed to an $Fe^{2+}$ --> $Fe^{3+}$ charge transfer transition in oxidized iron in phyllosilicates. The existence of aqueous alteration of near-subsurface material has been inferred for many asteroids based on the spectrophotometric evidence of phyllosilicate and iron alteration minerals (Rivkin et al., 2002).

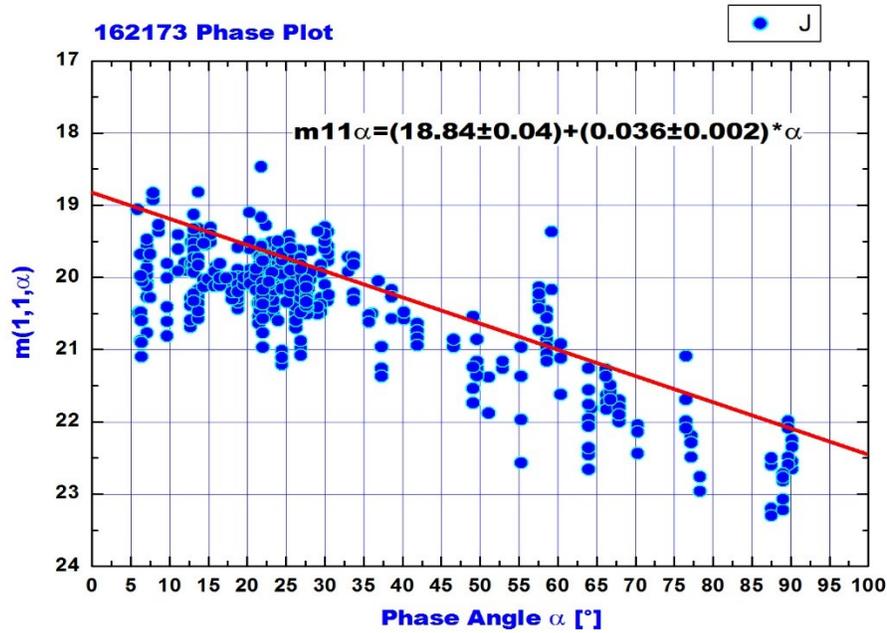

**Figure 12.** The phase plot of 162173 shows that a linear fit describes quite well the behavior of the comet. From this plot we find a β=0.036±0.002 to be compared with the result of Ishiguro et al. (2014) β= 0.039±0.001. The agreement is very satisfactory.

o o o o o o o o o o o o o o o

For 162173 Mueller et al. (2011) find a thermal inertia of the top layer of $150 - 300$ Jm$^{-2}$ s$^{-1/2}$ K$^{-1}$ . This is much larger than the thermal inertia of 9P/Tempel 1 ( $< 50$ Jm$^{-2}$ s$^{-1/2}$ K$^{-1}$). This result make sense if 162173 is much older than 9P, and as a consequence, the depth of the top dust layer were much deeper than in 9P.

Busarev et al. (2017) made a study of 162173 and concluded that spectra of Ryugu taken on 2007 July 11th by Vilas (2008) and on 2012 July 5[th] by Sugita et al. (2013) show evidence of scattering light reflected from the surface in the asteroid's coma, formed by sublimed ice particles. The first spectrum was taken after aphelion and the second before aphelion. Their location has been plotted on our SLC (Figure 13) and agree with the idea of activity at aphelion. This activity might be explained if one rotational pole of the object is pointing to the Sun at aphelion, like in the case of comet 2P/Encke who shows extensive activity at aphelion (Ferrín, 2008). Busarev et al. (2017) conclude that the presumed temporal sublimation and dust activity is an indication of existence of a residual frozen core which could point to a recent transition from the main belt to the near-Earth region.

### 3.4.1 Our results

The phase plot of 162173 Ryugu is shown in Figure 12 and the SLC in Figure 13, exhibiting low level cometary activity. Interestingly the activity is near aphelion which may

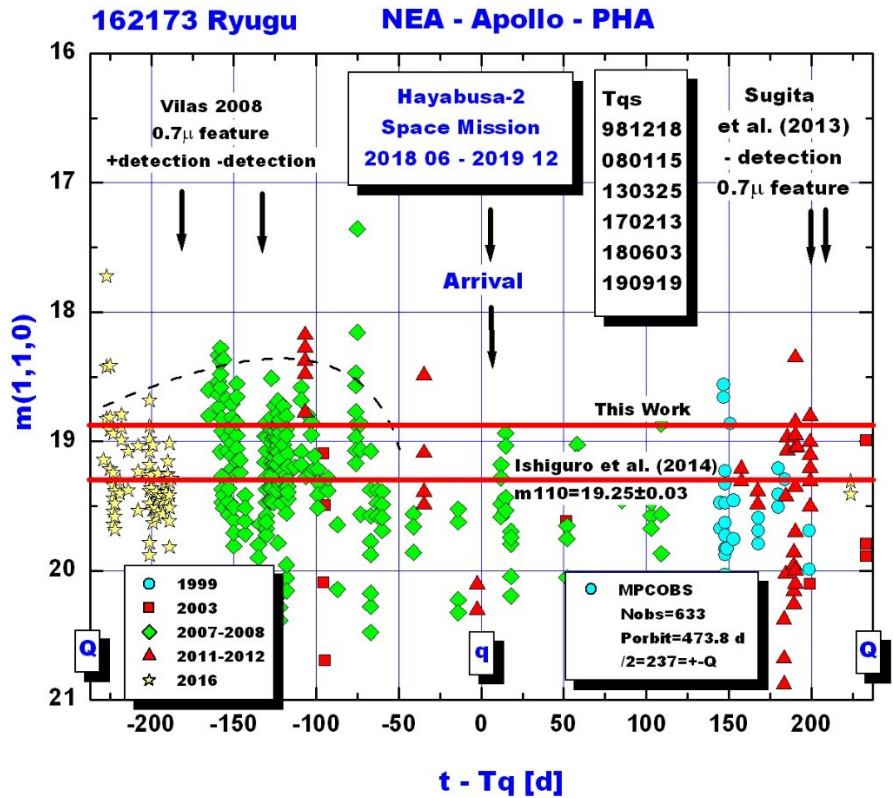

**Figure 13.** 162173 Ryugu is a NEA – Apollo – PHA target of the Hayabusa 2 spacecraft mission. The SLC exhibits an enhancement or bump with $(A_{SEC}, \Delta t) = (\sim -0.75, \sim 200\ d)$, consistent with low level cometary activity far from perihelion. This result is supported by the classification (C), albedo $p_V = 0.04\text{-}0.05$ measured by Muller et al. (2016) and ratified by Ishiguro et al. (2014) $p_V = 0.047 \pm 0.03$, and location on the B-V vs V-R and the R-I vs V-R colour-colour diagrams (Figures 20 and 21), values characteristic of other comets. At the time of arrival of Hayabusa-2 the object is inactive, but the mission control might detect the onset of this activity later on. Busarev et al. (2017) argue that spectra taken by Vilas (2008) and Sugita et al. (2013) on the dates arrowed exhibit cometary features a result consistent with the idea that there is aphelion activity. The second date by Vilas (2008) might have been negative because the data shows a deep reminiscent of an eclipse, implying the possibility that Ryugu might be a binary. This feature is enlarged in the next Figure. Recurrent activity at different returns is currently considered the most reliable indicator of sublimation driven activity (Hsieh et al., 2015). Liang-liang and Jiang-hui (2014) measured an effective diameter $D_{EFFEC} = 1.13 \pm 0.03$ km and an albedo $p_V = 0.042 \pm 0.003$, which corresponds to an absolute magnitude $m_V(1,1,0) = 18.82 \pm 0.14$. In this work (Table 1) and in this plot we find $m_V(1,1,0) = 18.84 \pm 0.04$, in perfect agreement.

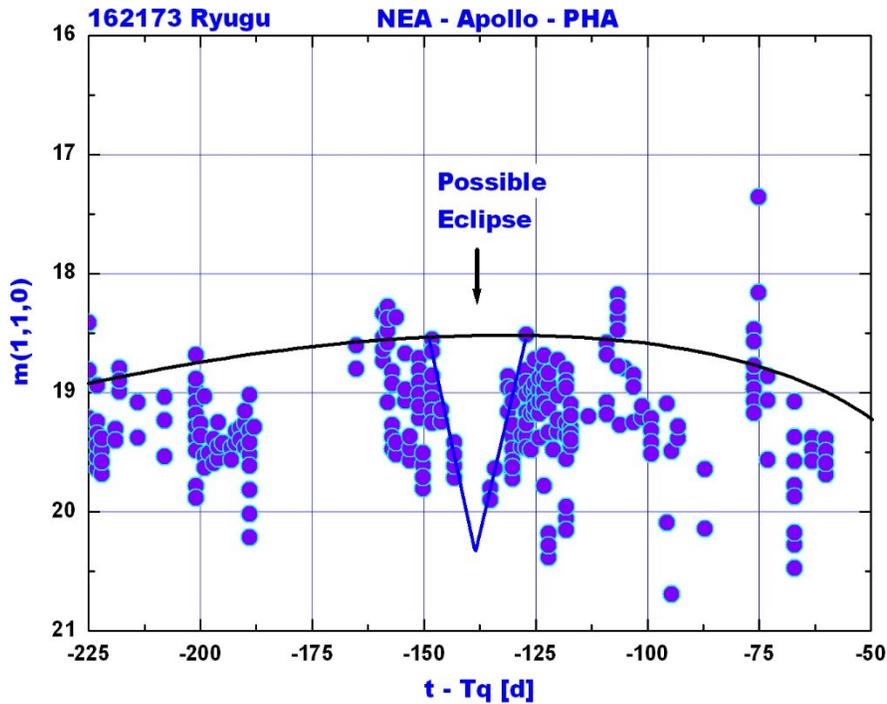

**Figure 14.** The eclipse feature of 162173 is enlarged in this Figure. There is a non-zero small probability that this feature is an statistical fluke, however in case it were not, it would imply that Ryugu might be a binary. One possible interpretation of the eclipse of depth $\Delta m = 1.7 \pm 0.2$ mag and duration $21.4 \pm 2.2$ days, is that the ratio of radii of the primary to the secondary is $R_P/r_S = \sim 1.9$. However the data is not sufficient and this interpretation is not unique or accurate. The next time when this region will be observable from ground based telescopes is starting on 2019 07 05 with Elongation=106° and magnitude m= 20.9. Observations are encouraged.

0 0 0 0 0 0 0 0 0 0 0 0 0 0 0

indicate that a pole of the asteroid containing water points to the Sun at - 200 < t-Tq < -100 d not far from aphelion. Sugita et al. (2013) found evidence that the pole of the asteroid is highly tilted, concluding that it probably pointed to the Sun at the end June of 2012 or ~-270 d before perihelion and near aphelion, supporting our conclusion.

In Figure 13 a signature in the shape of a V is clearly discerned. This might be an eclipse. The eclipse region is enlarged in Figure 14. The data is scarce and we can not determine if the depth of the eclipse is flat or in the shape of a V. We can only estimate that the ratio of radii of the primary to the secondary is $r_P/r_s \sim 1.9$. Thus might be 162173 is a binary.

The colour values derived above are plotted in Figures 20 and 21, and lie among other comets.

Ferrín (2014) defined age of a comet using as a proxy the inverse of the total mass loss per orbit. Using this definition the mass-loss-age for comet 9P/Tempel 1 and 162173 are 28 and ~4040 cy (comet years). Thus 162173 is much older than 9P. In fact 162173 is the third oldest comet in the database after 289P/Blanpain (~26.000 cy) and 107P/Wilson-Harrinton (~17.000 cy). Thus the AXA team of the Haybusa-2 space mission might have selected a very old low activity C-class comet rather than a C-class asteroid.

The location of 162173 in the $A_{SEC}$ vs Diameter diagram is shown in Figure 22. The object lies inside the graveyard defined as $A_{SEC}$ <1.0 magnitudes.

## 3.5 495848 = 2002 QD7

Asteroid 495848 is a NEA an Apollo and a PHA. Apart from the Minor Planet Center and JPL Minor Body Node, there is very little physical information on this object.

Observations of this object were carried out with the 1 m f/5 reflecting telescope of the National Observatory of Venezuela, CIDA. A FLI PL4240 CCD of 1075x1075 pixels with a V filter was used for the observations. Standard reduction and photometry were made using Maxim-DL which is more practical than IRAF for reducing comet images. We used a measuring aperture 5xFWHM in order to include any possible coma, however none was found. Photometric standards were taken from APASS catalog of the AAVSO (AAVSO Photometric All-Sky Survey). The log of observations is presented in Table 2. We routinely achieve a photometric error of ±0.01 mag = ±1% after averaging ~20 images. Figure 15 shows the average images on dates 2018 01 06 and 07, and the average of all 137 images with a total exposure time of 548 minutes = 9.13 hours. The object looks stellar in the image, and this is confirmed in a comparison of the FWHM of asteroid and star in the lower right hand image of Figure 15.

If the object is nuclear then we can use the mean magnitude to estimate the diameter of the object. From Table 2 we find a <m (1,1,0)>=18.32±0.03 which is fainter than the absolute magnitude given by the MPC by +0.32 mag. If we assume a geometric visual albedo pv=0.04 then the diameter is D=1.43±0.10 km.

The phase plot shown in Figure 16 shows a slope value β=0.031± 0.002, while the SLC shown in Figure 17 shows an enhancement in 2002-2007 and 2013-2017 centred at perihelion, but no activity in 2018. Perhaps our observing campaign started too late when the object was no longer sublimating. Although we tentatively conclude that a positive detection was manifest, observations should continue.

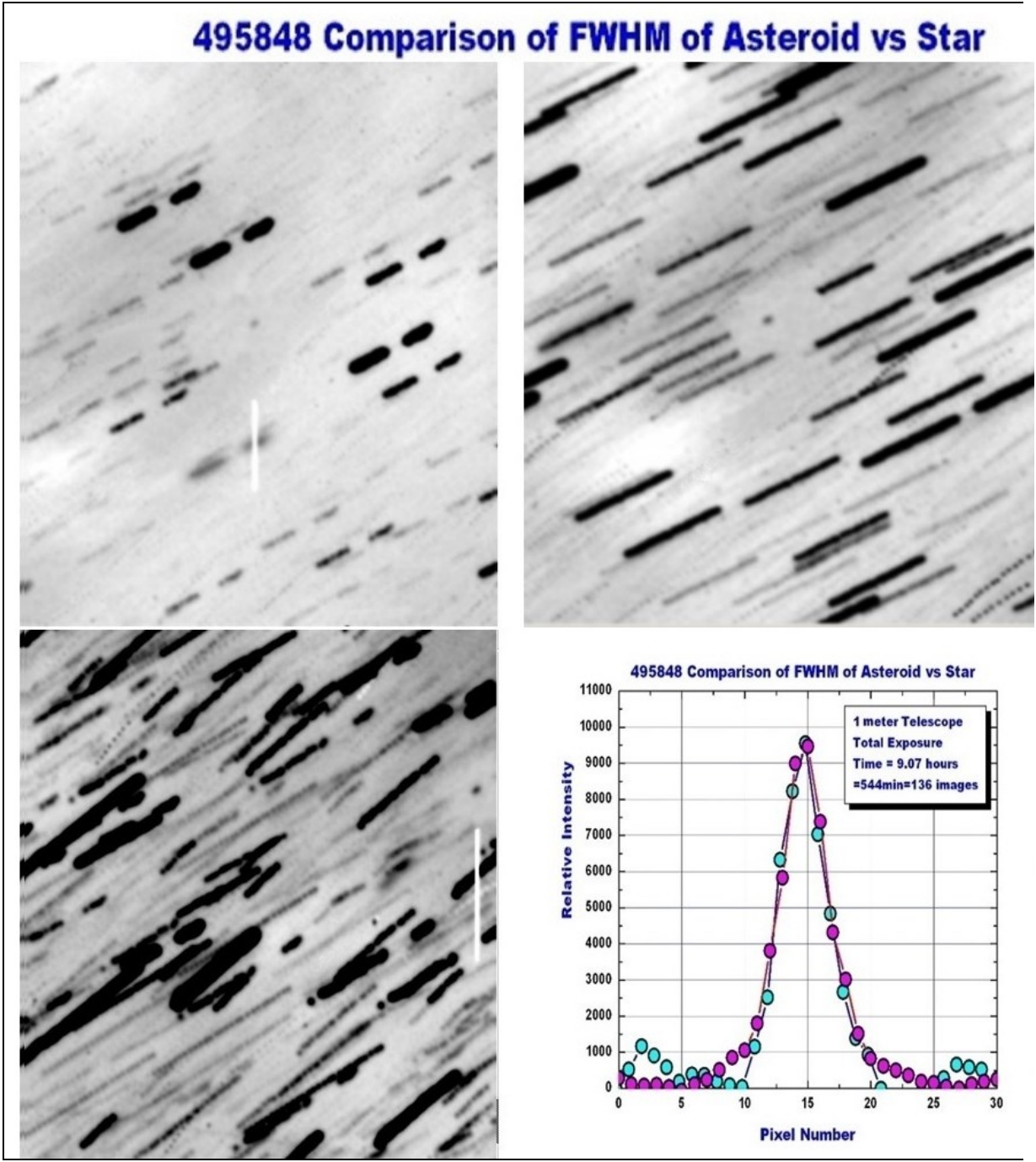

**Figure 15.** Images of 495848 taken with the 1 reflecting telescope of the National Observatory of Venezuela, CIDA. Upper left, 2018 January 6[th], 64 min of total exposure time. Upper right, 2018 January 7[th], 72 minutes of exposure time. Lower left, 564 min = 9.4 hours of total exposure time. The object is near the right border. Lower right, the profiles of asteroid and star were taken perpendicular to the direction of motion, and are compared in this plot. The asteroid has a star like profile, suggesting that it did not show cometary activity during the observing window. Thus it is feasible to derive the nuclear magnitude and diameter. We find m(1,1,0)=18.32±0.03, diameter D=1.43±0.10 km assuming $p_V$=0.04. The images are 8.6'x8.6' minutes of arc.

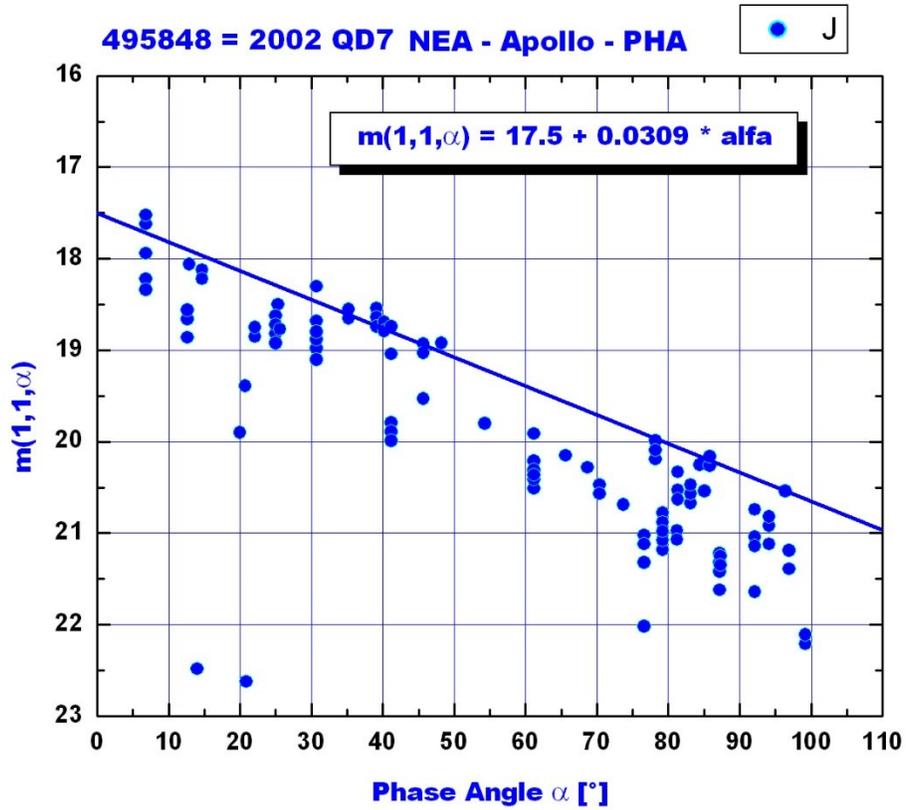

**Figure 16.** Phase plot of 495848. The distribution is linear up to phase angle 100°. All data with <5.5° have been eliminated from the plot, thus the opposition effect can not influence this SLC or produce false positives.

0 0 0 0 0 0 0 0 0 0 0 0

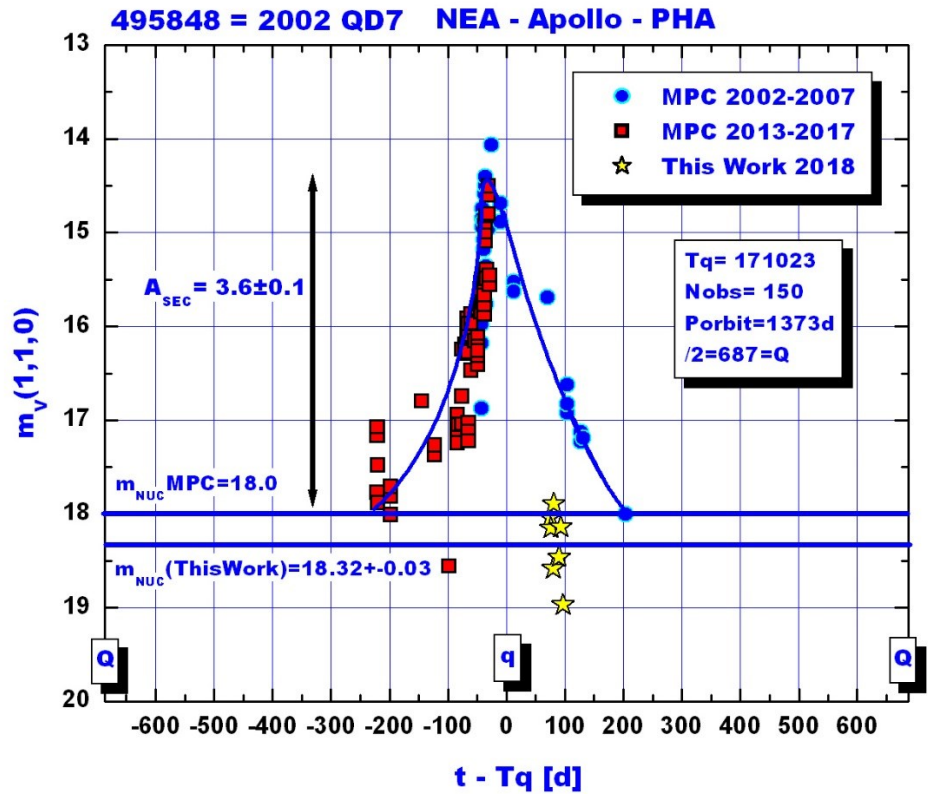

**Figure 17.** The SLC of 495848 shows evidence of cometary activity centred at perihelion in two oppositions, leaving little doubt that this a comet of significant activity. For this object we find $(A_{SEC}, \Delta t) = (-3.6, 550d)$. The stars are our observations taken with the 1m telescope of the National Observatory of Venezuela when the object was starlike. Thus it is possible to derive a nucleus magnitude and assuming a geometric albedo calculate a diameter. This is done in the text. Recurrent activity at different returns is currently considered the most reliable indicator of sublimation driven activity (Hsieh et al., 2015).

0 0 0 0 0 0 0 0 0 0 0 0

### 3.6 6063 Jason

The MPC data base list this object with an absolute magnitude H=15.9 and perihelion distance 0.51 AU. Thus again observation inside the Earth's orbit are difficult or impossible. The CALL database (Warner et al., 2009) gives a diameter of 1.96 km, an albedo of 0.20 and a rotational period of 51.7 h.

Figure 18 shows the phase plot and Figure 19 the SLC exhibiting low level cometary activity centered at perihelion. Figure 22 shows the location of 6063 on the $A_{SEC}$ vs Diameter plot. The object is located inside the graveyard.

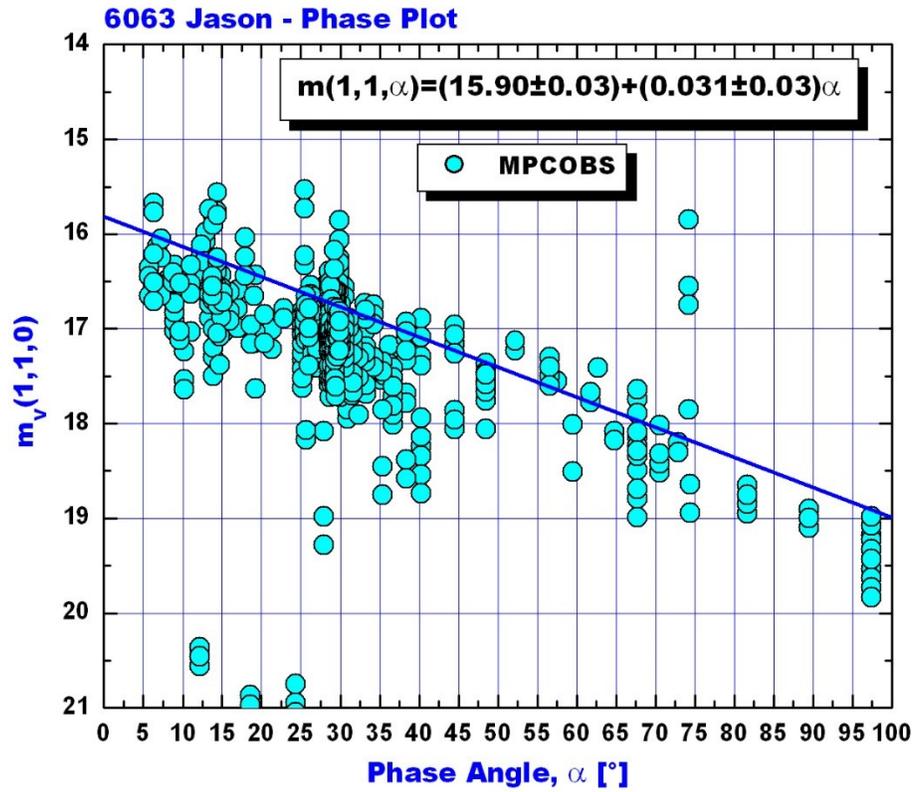

**Figure 18.** The phase plot of 6063 Jason shown a typical distribution with m(1,1,α) = (15.90±0.03) + (0.031±0.03) α.

0 0 0 0 0 0 0 0 0 0 0 0 0 0 0

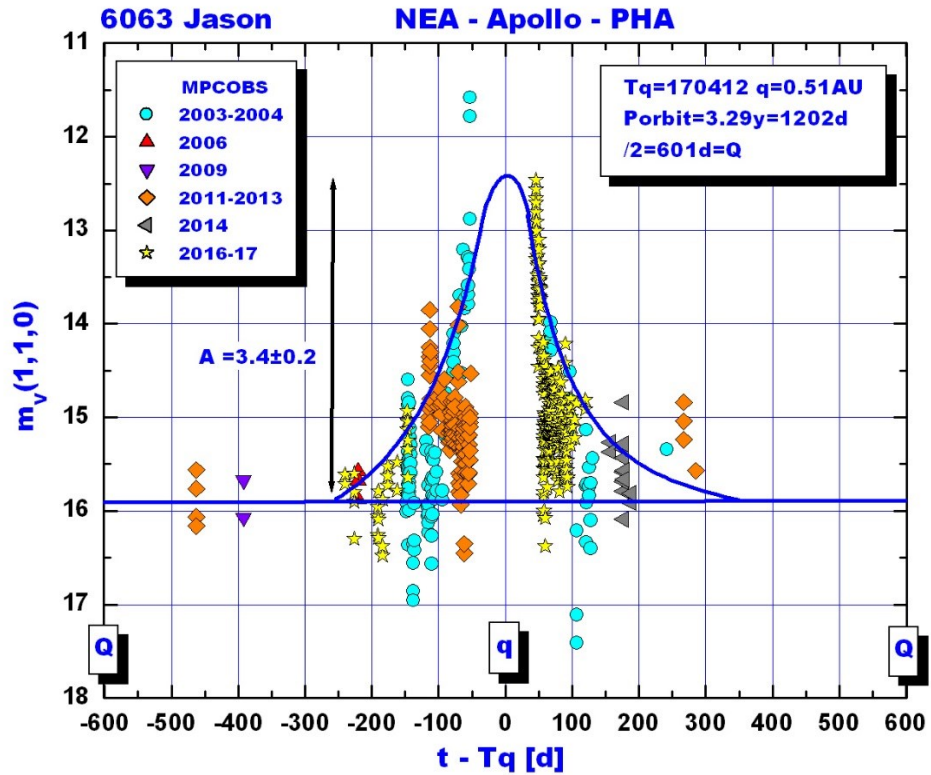

**Figure 19.** The SLC of 6063 Jason shows cometary activity centered at perihelion. For this case we find $(A_{SEC}, \Delta t) = (\sim{-}3.4, \sim{550}d)$. Recurrent activity at different returns is currently considered the most reliable indicator of sublimation driven activity (Hsieh et al., 2015).

0 0 0 0 0 0 0 0 0 0 0 0 0 0 0 0

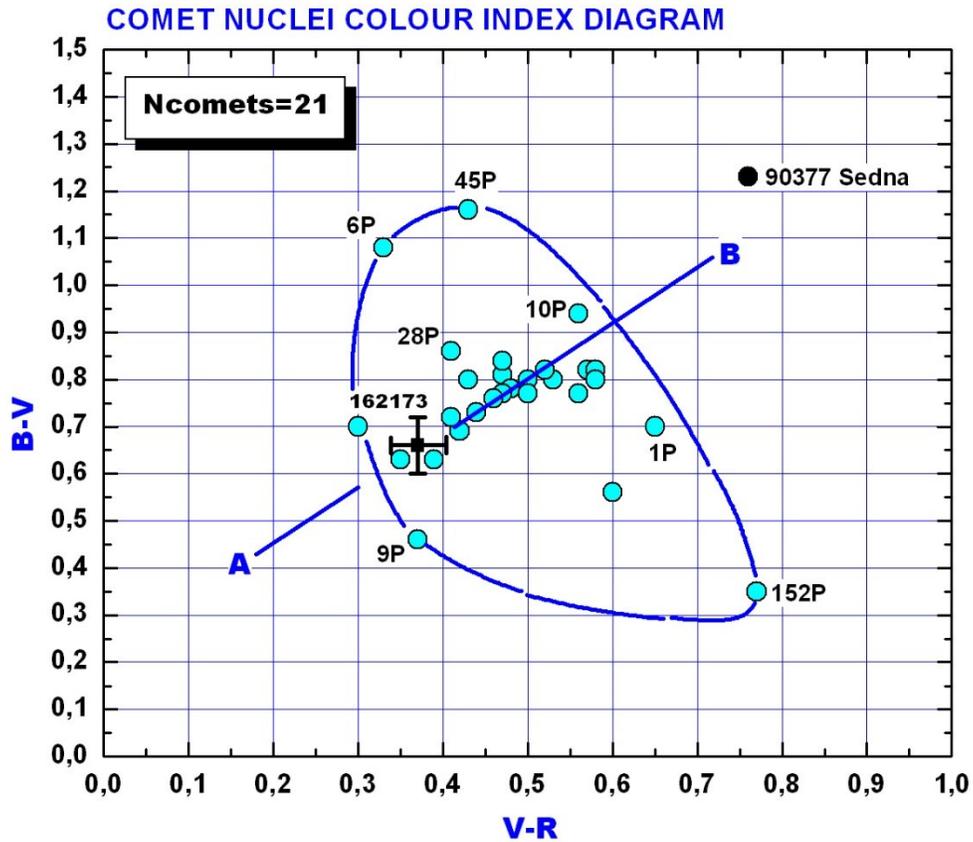

**Figure 20.** The B-V vs V-R colour diagram supports the conclusion that 162173 has colours consistent with those of comets, and in fact they lie on the blue line, the "principal component" for cometary colours. The colours of 162173 have been determined by Ishiguro et al. (2014) and Kawakami (2009).

o o o o o o o o o o o o

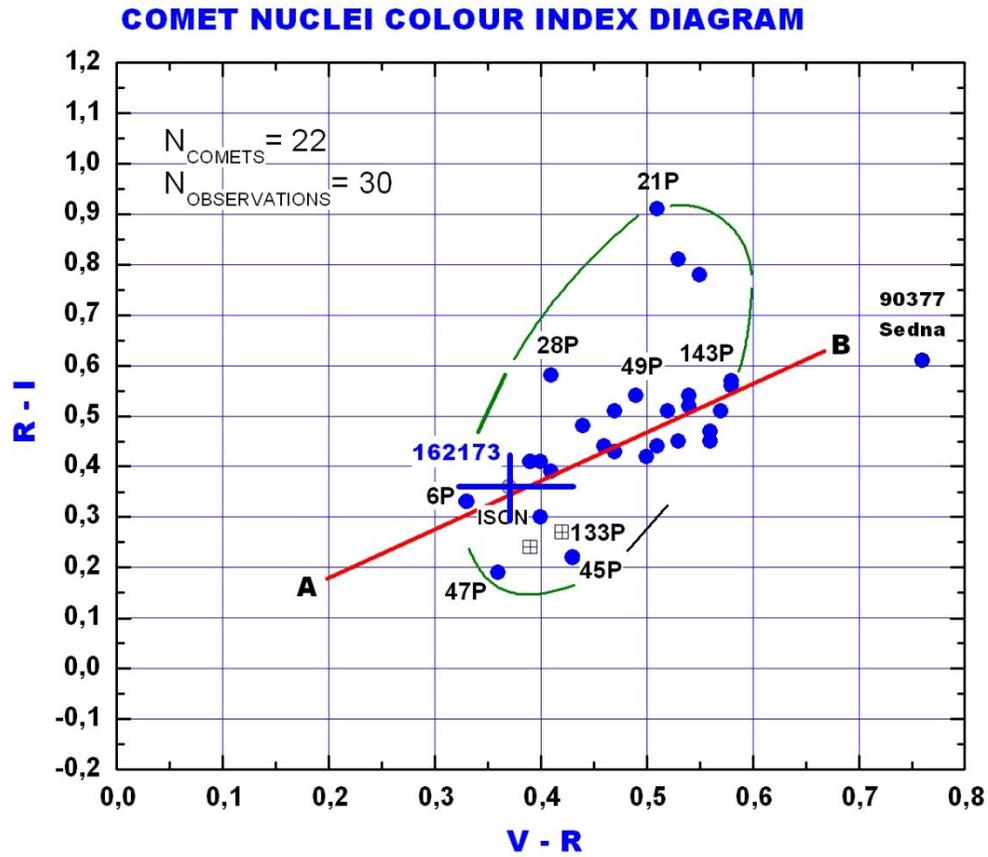

**Figure 21.** This R-I vs V-R colour diagram supports the conclusion that 162173 has the colour of comets, and in fact it lies on the red line A-B, which is the "principal component" for cometary colours. The colours of 162173 have been determined by Ishiguro et al. (2014) and Kawakami (2009).

o o o o o o o o o o o

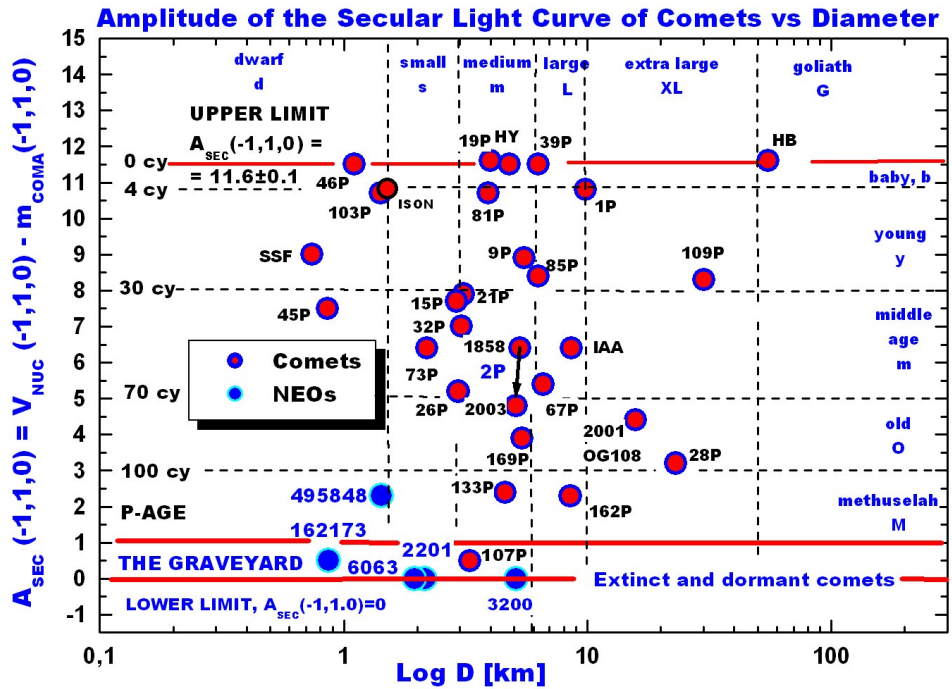

**Figure 22.** Amplitude of the secular light curve, $A_{SEC}$, vs effective diameter D (Ferrín, 2014). For the asteroids $A_{SEC}$ was converted from R=q to R=1 AU before plotting. Three object lie in the graveyard defined as $A_{SEC}$ <1.0 magnitude: 162173, 6063 and 2201. Thus all of them are old objects, methuselah comets, compared with other younger comets. The other object that lies in the graveyard is 107P/Wilson-Harrington. The photometric age, P-AGE, is an attempt to define an age for comets using as a proxy the amplitude of the SLC. Notice that in the diagram five objects have the same upper limit $A_{SEC}$ =11.6±0.01, which means that there is an upper limit to the surface activity of a comet. This information can be used to set lower size limits for objects for which the diameter is unknown. One significant conclusion from this plot is that most NEAs lie at the bottom, near or inside the comets´ graveyard.

o o o o o o o o o o   o o o

## 4. Conclusions

Using the Secular Light Curve (SLC) formalism, we searched for cometary activity of six NEAS finding positive evidence in four cases. We also determined their phase curves and in one case the size of the object.

**(1) 2201 Oljato.** This asteroid has been suspected of cometary activity since 1963. The SLC shows a non-flat distribution with maximum intensity at perihelion, suggesting significant cometary activity. This enhancement is repeated at many oppositions and is characteristic of many other comets. Additionally the SLC shows evidence of an eclipse.

**(2) 3200 Phaethon.** Observed activity near perihelion using STEREO observations is not sufficient to supply the associated meteor stream mass. The SLC shows a flat distribution and no evidence of cometary activity. This is rather surprising. We conclude that 3200 is a dormant comet that was active in the past and now has ceased its activity. The possibility of being an extinct comet has been ruled out.

**(3) 99942 Apophis.** This asteroid will have a very close encounter with Earth on April 13[th], 2029, so it in interesting to search for possible cometary activity. The SLC does not show evidence of activity but there is evidence of an eclipse suggesting that 99942 might be a contact binary asteroid. This is confirmed using Goldstone and Arecibo radar data, and an image is provided.

**(4) 162173 Ryugu.** This NEA is the target of the Hayabusa 2 spacecraft mission. The evidence for 162173 being a comet is positive. The SLC exhibits an enhancement or bump with (Asec,$\Delta$t)=($\sim$-0.75,$\sim$125 d), where $A_{SEC}$ is the amplitude of the SLC and $\Delta$t the duration of the activity, consistent with cometary activity. This result is supported by the classification (C type), albedo (0.04-0.05), and location on the B-V vs V-R and R-I vs V-R colour-colour diagrams. Thus the JAXA team of the Haybusa-2 space mission might have selected an *old, low activity, C-class comet* rather than a C-class asteroid. They will be able to confirm or deny this activity near the end of the mission.

**(5) 495848.** Apart from the information given by the Minor Planet Center and the JPL Minor Body Node, there is very little physical information on this object. The object was observed from the National Observatory of Venezuela in 7 nights during which the asteroid was stellar-like. Thus it was possible to derive a mean nuclear magnitude $<m_{NUC} (1,1,0)>$=18.32$\pm$0.03 and if a geometric albedo is assumed, $p_V$=0.04, then a nuclear diameter can be calculated, D=1.43$\pm$0.10 km. The SLC shows evidence of cometary activity.

**(6) 6063 Jason.** The SLC of this NEA also exhibits positive cometary activity centered at perihelion.

**(7)** One significant conclusion from this investigation is that in the $A_{SEC}$ vs Diameter plot (Figure 22), all positive NEAs lie at the bottom of the diagram, and three of them inside the comets´ graveyard.

**(8)** These results show that the SLC formalism is a useful method to detect low level cometary activity in objects that do not exhibit a coma or a tail, and that currently it is the only method capable of doing so. Additionally it has been shown that the accuracy of the MPC data is sufficient to detect mutual eclipses in those objects, thus detecting for the first time or confirming binarity on those objects.

**(9)** Although in this work we have not done an exhaustive and profound search of the NEAs database, but nevertheless we were able to detect four positive objects among its members, hints that the database may hide many more low-level active objects as yet undetected.

## 5. Acknowledgements

The FACom group is supported by the project "Estrategia de Sostenibilidad 2015 - 2016", sponsored by the Vicerectoría de Investigación of the Universidad de Antioquia, Medellín, Colombia. This work contains observations made at the National Observatory of Venezuela, Centro de Investigaciones de Astronomía, CIDA, Mérida. The help of Giuliatna Navas and the night assistants Richard Rojas, Leandro Araque, Dalbare González, Fredy Moreno, Carlos Pérez, Gregory Rojas, Daniel Cardozo and Ubaldo Sánchez, is particularly appreciated. We used as comparison stars the APASS catalog published by the AAVSO. We acknowledge with thanks the comet observations from the COBS Comet Observation Database, contributed by observers worldwide and used in this research. We thank an anonymous referee for scientific suggestions that improved the quality of this manuscript significantly.

Table 1.  Numerical Results

| NEA[a] | Family Earth MOID[AU][b] | $m_V(1,1,0)$[c] | $\beta$[d] | $H_V$[e] | $G$[e] | $A_{SEC}(q)$ $A_{SEC}(1)$[f] | $\Delta t$[g] [d] |
|---|---|---|---|---|---|---|---|
| 2201 | Apollo, PHA 0.00314 | 15.3±0.1 | 0.034 ±0.002 | 15.25 | 0.15 | ~2.1 ~0.0 | ~700 |
| 3200 | Apollo, PHA 0.01989 | 16.67±0.03 | 0.035 ±0.003 | 14.2 | 0.15 | - - - - - - - - | 0 |
| 99942 | Aten->Apollo, PHA 0.00017 | 18.0±0.1 | 0.043 ±0.003 | 19.2 | 0.15 | - - - - - - - - | 0 |
| 162173 | Apollo, PHA 0.00012 | 18.84±0.04 | 0.036 ±0.002 | 19.3 | 0.15 | ~0.60 ~0.5 | ~350 |
| 495848 | Apollo, PHA 0.10 | 18.3±0.2 | 0.031 ±0.002 | 18.0 | 0.15 | ~3.6 ~2.3 | ~550 |
| 6063 | Apollo 0.07444 | 15.90±0.03 | 0.032 ±0.003 | 15.9 | 0.15 | ~3.4 ~0.0 | ~550 |

a- Asteroid number,  b- Family and Earth's MOID,  c- absolute magnitude as defined in this work,  d- phase coefficient determined in this work,  e- Hv and G values from Veres et al. (2015),  f- Amplitude of the Secular Light Curve at R=q and R= 1AU,  g- duration of activity in days. Notice that the activity of 2201 and 6063 is entirely due to their small perihelion distance and that if they were displaced to q=1.0 AU their activity would be cero.

Table 2. Log of observations of 495848 with the 1m reflector telescope of the National Observatory of Venezuela, CIDA:

| Date | $\Delta$[AU] | R[AU] | E[°] | $\alpha$[°] | # of imag | TET [min] | V Observed | $m_V(1,1,0)$ | t-Tq [d] |
|---|---|---|---|---|---|---|---|---|---|
| 2018 01 06 | 0.533 | 1.345 | 122.2 | 38.2 | 12 | 48 | 18.50±0.02 | 18.04±0.02 | +75 |
| 2018 01 07 | 0.536 | 1.356 | 123.7 | 37.1 | 18 | 72 | 18.60±0.04 | 18.15±0.04 | +76 |
| 2018 01 11 | 0.549 | 1.394 | 128.7 | 33.4 | 27 | 108 | 19.03±0.01 | 18.58±0.01 | +80 |
| 2018 01 12 | 0.552 | 1.403 | 129.9 | 32.5 | 24 | 96 | 18.34±0.01 | 17.89±0.01 | +81 |
| 2018 01 21 | 0.587 | 1.486 | 140.8 | 24.7 | 29 | 116 | 18.93±0.01 | 18.46±0.01 | +90 |
| 2018 01 24 | 0.602 | 1.515 | 144.3 | 22.3 | 18 | 72 | 18.63±0.01 | 18.14±0.01 | +93 |
| 2018 02 28 | 0.624 | 1.552 | 149.4 | 18.7 | 9 | 36 | 19.50±0.01 | 18.97±0.01 | +97 |
| Totals | | | | | 137 | 548 | | <18.32±.03> | |

- The columns give the following information: The date, Earth distance, Sun distance, Elongation, phase angle, number of images taken, TET = total exposure time, observed V magnitude, absolute magnitude $m_V(1,1,0)$ and time with respect to perihelion.  The mean absolute magnitude during the observed window is <V>=18.32±0.03 which is used in the text to calculate the diameter of the object.